\documentclass[twocolumn,showpacs,preprintnumbers,amsmath,amssymb]{revtex4}

\usepackage{graphicx}
\usepackage{dcolumn}
\usepackage{bm}

\RequirePackage{cmap}
\RequirePackage[cp1251]{inputenc}
\RequirePackage[TS1,T2A]{fontenc}

\newcommand{\frat}[2]{\frac{\textstyle #1}{\textstyle #2}}

\begin{document}

\title{Thermodynamics of quark quasi-particle ensemble}

\author{S. V. Molodtsov}
 \altaffiliation[Also at ]{
Institute of Theoretical and Experimental Physics, Moscow, RUSSIA}
\affiliation{%
Joint Institute for Nuclear Research, Dubna,
Moscow region, RUSSIA
}%
\author{G. M. Zinovjev}
\affiliation{
Bogolyubov Institute for Theoretical Physics,
National Academy of Sciences of Ukraine, Kiev, UKRAINE
}%

\date{\today}

\date{\today}

\begin{abstract}
The features of hot and dense gas of quarks which are considered as the
quasi-particles of the model Hamiltonian with four-fermion interaction
are studied. Being adapted to the Nambu-Jona-Lasinio model this approach
allows us to accommodate a phase transition similar to the nuclear
liquid-gas one at the proper scale and to argue an existence of the mixed
(inhomogeneous) phase of vacuum and normal baryonic matter as a plausible
scenario of chiral symmetry (partial) restoration. Analyzing the transition
layer between two phases we estimate the surface tension coefficient and
speculate on the possible existence of quark droplet.
\end{abstract}

\pacs{11.10.-z, 11.15.Tk}     
\maketitle

Notwithstanding the well-known incompletion of quantum chromodynamics (QCD)
and its strongly limited capacity to perform the non-perturbative calculations
both in the vacuum and at the finite temperature and baryonic density, one of
the major predictions of this theory, a possible existence of quark-gluon
plasma, was so exciting and convincing that pushed forward very active
research programme in experiments with relativistic
heavy ions. In such a situation when the applications which should usually be
based on the standard interrelations between the hadronic properties and the
QCD Lagrangian parameters are merely impossible, but the practical need in the
quantitative estimates is dictated by the running experiments, we are forced to be
very persistent and pragmatic in searching the effective Lagrangians.

The Nambu-Jona-Lasinio (NJL) model and its numerous extensions are the
most popular in this context because they share some global symmetries of QCD
and allow us to make surmountable some serious difficulties and uncertainties
faced in the QCD calculations. It appears especially appreciable and effective
in studying the nature of nuclear matter and its (super)dense state being
treated as a model of QCD at large quark chemical potential. Nowadays the
experiments in heavy-ion collisions to a considerable extent are driven by the results
of phenomenological investigations of the properties of nucleon-nucleon force
and the phase diagram of strongly interacting matter which relies on the
corresponding estimates of experimentally measurable quantities.

Thus, exploring the QCD phase diagram with the effective models is targeted
from the theoretical view point by necessity to find out some kind of
interpolation between the physics as conceived by the lattice QCD simulations
(still unrealistic because of several reasons) and the physics outputs of
phenomenological studies. The vast activity (and progress) along this line
\cite{Wamb} together with the recent experimental results from LHC and RHIC
\cite{Borg} lead to the unexpected vision of still pending questions and,
perhaps, their new interpretation.

These and some related topics are discussed in this paper inspired by well known
and fruitful idea about the specific role of surface degrees of freedom in the
finite fermi-liquid systems \cite{Kh} and, to a considerable extent, by our
previous works \cite{MZ}, \cite{ff} in which the quarks were treated as the
quasi-particles of the model Hamiltonian, and the problem of filling up the
Fermi sphere was studied in detail. In particular, within the Nambu-Jona-Lasinio
(NJL) model \cite{4} new solution branches of the equation for dynamical quark
mass as a function of chemical potential (the details are shown below in
Fig. \ref{f1}) have been found out. Besides, the existence and origin of the
state filled up with quarks which is almost degenerate with the vacuum state
both in the quasi-particle chemical potential and in the ensemble pressure has been
demonstrated. In general, the approach developed may be considered as another
microscopical substantiation of the bag model in which the states filled up with
quarks might be instrumental as a 'construction material' for baryons.

Our analysis here is performed within two approaches which are supplementary, in
a sense, but, fortunately, lead to the identical results. One of these approaches based on
the Bogolyubov transformation is especially informative for studying the process
of filling the Fermi sphere up because the density of quark ensemble develops a
continuous dependence on the Fermi momentum in this case. It allows us to reveal
an additional structure in the solution of gap equation for dynamical quark mass
just in the proper interval of parameters characteristic for the phase
transition and to trace its evolution. It results in the possibility for quark (fermionic)
ensemble to be found in two aggregate states, a gas and a liquid, and the
partial restoration of chiral condensate in a liquid phase (Section I). In order to make
these conclusions easily perceptible we deal with the simplest version of the
NJL model (with one flavor and one of the standard parameter sets). We try also to
construct a description of transition layer between two phases and, in
particular, to estimate the surface tension coefficient (Section II) what is of obvious
importance in context of discussing the possible quark droplet formation
(Section III). Some technical moments of calculating the mean energy functional are picked out
into Appendix.

\section{Exploring quark ensemble}
Now as an input for starting we remind the key elements of approach which has
been developed  \cite{MZ}, \cite{ff}. The corresponding model Hamiltonian
includes the interaction term taken in the form of a product of two colour
currents located in the spatial points
${\bf x}$ and ${\bf y}$ which are connected by a form-factor and its density reads as
\begin{equation}
\label{1}
{\cal H}=-\bar q(i{\bf \gamma}{\bf \nabla}+im)q-\bar q t^a\gamma_\mu q
\int d{\bf y}\bar q' t^b\gamma_\nu q' \langle A^{a}_\mu A'^{b}_\nu\rangle,
\end{equation}
where $q=q({\bf x})$, $\bar q=\bar q({\bf x})$, $q'=q({\bf y})$,
$\bar q'=\bar q({\bf y})$ are the quark and anti-quark operators,
\begin{eqnarray}
\label{2}
&&\hspace{-0.5cm}
q_{\alpha i}({\bf x})=\int\frac{d {\bf p}}{(2\pi)^3}
\frac{1}{(2|p_4|)^{1/2}}
\left[a({\bf p},s,c)u_{\alpha i}({\bf p},s,c) e^{i{\bf p}{\bf x}}
+\right.\nonumber\\[-.2cm]
\\ [-.25cm]
&&~~~~\left.+b^+({\bf p},s,c)v_{\alpha i}({\bf p},s,c) e^{-i{\bf p}{\bf
x}}\right],
\nonumber
\end{eqnarray}
$p_4^2=-{\bf p}^2-m^2$, $i$--is the colour index, $\alpha$ is the spinor index
in the coordinate space, $a^+$, $a$ and $b^+$, $b$ are the creation and annihilation
operators of quarks and anti-quarks, $a~|0\rangle=0$, $b~|0\rangle=0$, $|0\rangle$ is the
vacuum state of free Hamiltonian and $m$ is a current quark mass. The summation over
indices $s$ and $c$ is meant everywhere, the index $s$ describes two spin polarizations of
quark and the index $c$ plays the similar role for a colour. As usual
$t^a=\lambda^a/2$ are the generators of $SU(N_c)$ colour gauge group. The Hamiltonian density is
considered in the Euclidean space and $\gamma_\mu$ denote the Hermitian Dirac matrices,
$\mu,\nu=1,2,3,4$. $\langle A^{a}_\mu A'^{b}_\nu\rangle$ stands for the form-factor of the
following form
\begin{equation}
\label{cor}
\langle A^{a}_\mu A'^{b}_\nu\rangle=\delta^{ab}
 \frac{2~\widetilde G}{N_c^2-1}\left[I({\bf x}-{\bf y})
\delta_{\mu\nu}-J_{\mu\nu}({\bf x}-{\bf y})\right],
\end{equation}
where the second term is spanned by the relative distance vector and the gluon
field primed denotes that in the spatial point ${\bf y}$. The effective Hamiltonian density
(\ref{1}) results from averaging the ensemble of quarks influenced by intensive stochastic
gluon field $A^a_\mu$, see Ref. \cite{MZ}. For the sake of simplicity we neglect the
contribution of the second term in (\ref{cor}) in what follows. The ground state of the system is
searched as the Bogolyubov trial function composed by the quark-anti-quark pairs with opposite
momenta and with vacuum quantum numbers, i.e.
\begin{eqnarray}
\label{4}
&&\hspace{-0.65cm}|\sigma\rangle={\cal{T}}~|0\rangle~,~~~\nonumber\\[-
.2cm]
\\ [-.25cm]
&&\hspace{-0.65cm}
{\cal{T}}=\Pi_{ p,s}\exp\{\varphi[a^+({\bf p},s)b^+(-{\bf p},s)+
a({\bf p},s)b(-{\bf p},s)]\}.\nonumber
\end{eqnarray}
In this formula and below, in order to simplify the notations, we refer to one
compound index only which means both the spin and colour polarizations. The parameter
$\varphi({\bf p})$ which describes the pairing strength is determined by the minimum of mean energy
\begin{equation}
\label{5}
E=\langle\sigma|H|\sigma\rangle~.
\end{equation}
By introducing the 'dressing transformation' we define the creation and
annihilation operators of quasi-particles as $A={\cal{T}}~a~{\cal{T}}^{-1}$,
$B^+={\cal{T}}~b^+{\cal{T}}^{-1}$ and for fermions ${\cal{T}}^{-1}={\cal{T}}^\dagger$.
Then the quark field operators are presented as
\begin{eqnarray}
\label{6}
&&q({\bf x})=\int\frac{d {\bf p}}{(2\pi)^3} \frac{1}{(2|p_4|)^{1/2}}~
\left[~A({\bf p},s)~U({\bf p},s)~e^{i{\bf p}{\bf x}}+\right.\nonumber\\
&&~~~~~~~\left.+B^+({\bf p},s)~V({\bf p},s)~ e^{-i{\bf p}{\bf
x}}\right]~,\nonumber\\
&&\bar q({\bf x})=\int\frac{d {\bf p}}{(2\pi)^3}
\frac{1}{(2|p_4|)^{1/2}}~
\left[~A^+({\bf p},s)~\overline{U}({\bf p},s)~e^{-i{\bf p}{\bf
x}}+\right.\nonumber\\
&&~~~~~~~\left.+B({\bf p},s)~\overline{V}({\bf p},s)~ e^{i{\bf p}{\bf
x}}\right]~,
\nonumber
\end{eqnarray}
and the transformed spinors $U$ and $V$ are given by the following forms
\begin{eqnarray}
\label{7}
&&U({\bf p},s)=\cos(\varphi)~u({\bf p},s)-
\sin(\varphi)~v(-{\bf p},s)~,\nonumber\\[-.2cm]
\\ [-.25cm]
&&V({\bf p},s)=\sin(\varphi)~u(-{\bf p},s)+
\cos(\varphi)~v({\bf p},s)~.\nonumber
\end{eqnarray}
where $\overline{U}({\bf p},s)=U^+({\bf p},s)~\gamma_4$,
$\overline{V}({\bf p},s)=V^+({\bf p},s)~\gamma_4$ are the Dirac conjugated
spinors.

In Ref. \cite{ff} the process of filling in the Fermi sphere with the quasi-particles of
quarks was studied by constructing the state of the Sletter determinant type
\begin{equation}
\label{8}
|N\rangle=\prod_{|{\mbox{\scriptsize{\bf P}}}|<P_F;S}~A^+
({\bf P};S)~|\sigma\rangle~,
\end{equation}
which possesses the minimal mean energy over the state $|N\rangle$. The
polarization indices run through all permissible values here and the quark momenta
are bounded by the limiting Fermi momentum $P_F$. The momenta and polarizations
of states forming the quasi-particle gas are marked by the capital letters similar
to above formula and the small letters are used in all other cases.

As it is known the ensemble state at finite temperature $T$ is described by the
equilibrium statistical operator $\xi$. Here we use the Bogolyubov-Hartree-Fock
approximation in which the corresponding statistical operator is presented by the following form
\begin{equation}
\label{dm}
\xi=\frac{e^{-\beta ~\hat H_{{\mbox{\scriptsize{app}}}}}}{Z_0}~,
~~Z_0=\mbox{Tr}~\{e^{-\beta ~\hat H_{{\mbox{\scriptsize{app}}}}}\}~~,
\end{equation}
where an approximating effective Hamiltonian $H_{{\mbox{\scriptsize{app}}}}$ is
quadratic in the creation and annihilation operators of quark and anti-quark quasi-particles
$A^+$, $A$, $B^+$, $B$ and is defined in the corresponding Fock space with the vacuum state
$|\sigma\rangle$ and $\beta=T^{-1}$. There is no need to know the exact form of this operator
henceforth because all the quantities of our interest in the Bogolyubov-Hartree-Fock
approximation are expressed by the corresponding averages
$n(P)=\mbox{Tr} \{\xi A^+({\bf P};S) A({\bf P};S)\}$,
$\bar n(Q)=\mbox{Tr} \{\xi B^+({\bf Q};T) B({\bf Q};T)\}$
which are obtained by solving the following variational problem. The statistical
operator $\xi$ is defined by such a form in order to have the minimal value of
mean energy of quark ensemble
$$E=\mbox{Tr} \{\xi~H\}~$$
at the fixed mean charge
\begin{equation}
\label{ntot}
\bar Q_4=\mbox{Tr} \{\xi~Q_4\}=
V~2 N_c \int  \frac{d {\bf p}}{(2\pi)^3}~[n(p)-\bar n(p)]~,
\end{equation}
where
\begin{eqnarray}
Q_4&=&-\int~d{\bf x}~\bar q i \gamma_4 q=\nonumber\\
&=& \int\frac{d{\bf p}}{(2\pi)^3}\frac{-ip_4}{|p_4|}
\left[A^+(p)A(p)+B(p)B^+(p)\right]~,\nonumber
\end{eqnarray}
for the diagonal component (which is a point of our interest here) and at the
fixed mean entropy
($S=-\ln \xi$)
\begin{eqnarray}
\label{stot}
&&\hspace{-0.5cm}\bar S=-\mbox{Tr} \{\xi \ln \xi\}=\\
&&\hspace{-0.5cm}=-V 2 N_c \int\!\!\!  \frac{d {\bf p}}{(2\pi)^3}
\left[n(p)\ln n(p)+(1-n(p))\ln (1-n(p))+\right.\nonumber\\
&&\hspace{-0.5cm}\left.+\bar n(p)\ln \bar n(p)+(1-\bar n(p))
\ln (1-\bar n(p))\right].\nonumber
\end{eqnarray}
The mean charge (\ref{ntot}) is calculated here up to the unessential
(infinite) constant coming from permuting the operators $B B^+$ in the charge
operator $Q_4$. It is appropriate here to remind that the mean charge should
be treated in some statistical sense because it characterizes
quark ensemble density and has no colour indices.The mean energy density per one
quark degree of freedom $w={\cal E}/(2 N_c)$, ${\cal E}=E/V$ where $E$ is a total energy of
ensemble is calculated (the details of derivation can be found in Appendix)
to get the following form
\begin{widetext}
\begin{equation}
\label{15}
\hspace{-0.3cm}w=\int\frac{d {\bf p}}{(2\pi)^3} |p_4| +
\int\frac{d {\bf p}}{(2\pi)^3}|p_4|\cos\theta [n+\bar n-1]-
G\int \frac{d {\bf p}}{(2\pi)^3} \sin \left(\theta-\theta_m\right)
[n+\bar n-1]
\int \frac{d {\bf q}}{(2\pi)^3}\sin\left(\theta'-
\theta'_m\right)
[n'+\bar n'-1]I,
\end{equation}
\end{widetext}
where $\theta=2\varphi$, $\theta'=\theta(q)$, $n'=n(q)$, $I= I({\bf p}+{\bf
q})$ and the angle $\theta_m(p)$ is determined by $\sin \theta_m=m/|p_4|$.
We are interested in minimizing the following functional
\begin{equation}
\label{fun}
\Omega = E-\mu~\bar Q_4 -T~\bar S~,
\end{equation}
where $\mu$ and $T$ are the Lagrange factors for the chemical potential and
temperature respectively. The approximating Hamiltonian
$\hat H_{{\mbox{\scriptsize{app}}}}$ is constructed simply by using the
information on $E-\mu~\bar Q_4$ of presented functional
(see, also below). For the specific contribution per one quark degree of freedom
$f=F/(2N_c)$, $F=\Omega/V$ we receive
\begin{widetext}
\begin{eqnarray}
\label{17}
&&f=\int \frac{d{\bf p}}{(2\pi)^3}~
\left[|p_4|\cos\theta~(n+\bar n -1)-\mu~(n-\bar n)\right]
+\int \frac{d{\bf p}}{(2\pi)^3}~|p_4|-
G\int \frac{d{\bf p}}{(2\pi)^3}~
\sin \left(\theta-\theta_m\right)~(n+\bar n-1)\times\nonumber\\[-.2cm]
\\ [-.25cm]
&&\times
\int\frac{d{\bf q}}{(2\pi)^3}
~\sin\left(\theta'-\theta'_m\right)~(n'+\bar n'-1)~I+
T\int \frac{d{\bf p}}{(2\pi)^3}~
\left[n~\ln n+(1-n)~\ln(1-n)+\bar n~\ln \bar n+(1-\bar n)~
\ln(1-\bar n)\right]~.\nonumber
\end{eqnarray}
\end{widetext}
 The optimal values of parameters
are determined by solving the following equation system
($df/d\theta=0$, $df/d n=0$, $df/d \bar n=0$)
\begin{eqnarray}
\label{18}
&&|p_4|~\sin\theta-M\cos \left(\theta-\theta_m\right)=0~,\\
&&|p_4|~\cos\theta-\mu+M~\sin \left(\theta-\theta_m\right)-
T~\ln \left(n^{-1}-1\right)=0~,\nonumber\\
&&|p_4|~\cos\theta+\mu+M~\sin \left(\theta-\theta_m\right)-
T~\ln \left(\bar n^{-1}-1\right)=0~,\nonumber
\end{eqnarray}
where we denoted the induced quark mass as
\begin{equation}
\label{19}
\hspace{-0.1cm}
M({\bf p})=2G\!\!\!\int\!\!\! \frac{d{\bf q}}{(2\pi)^3}
(1-n'-\bar n')\sin \left(\theta'-\theta'_m\right)I({\bf p}+{\bf q}).
\end{equation}
\begin{figure}
\includegraphics[width=0.3\textwidth]{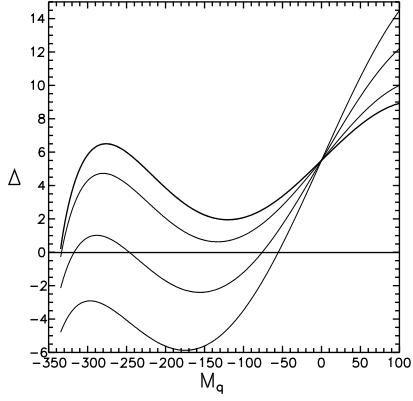}
\caption{The residual $\Delta$ for equation (\ref{19}) is presented as a
function of dynamical quark mass $M_q$ (MeV) at zero value of temperature
and the following values of chemical potential $\mu$ (MeV) --- $335$
(the lowest curve), $340$, $350$, $360$ (the top curve).}
\label{f1}
\end{figure}

Turning to the presentation of obtained results in the form customary for the
mean field approximation we introduce a dynamical quark mass $M_q$ parametrized as
\begin{equation}
\sin \left(\theta-\theta_m\right)=\frac{M_q}{|P_4|}~,~~
|P_4|=({\bf p}^2+M^2_q({\bf p}))^{1/2}~,
\end{equation}
and ascertain the interrelation between induced and dynamical quark masses. From
the first equation of system (\ref{18}) we fix the pairing angle
$$\sin \theta=\frac{p~M}{|p_4||P_4|}~$$ and making use the identity
\begin{equation}
\label{iden}
(|p_4|^2-M ~m)^2+M^2 p^2=[p^2+(M-m)^2]~|p_4|^2~,
\end{equation}
find out that $$\cos \theta=\pm\frac{|p_4|^2-m~M}{|p_4||P_4|}~.$$ For the sake
of clarity we choose the upper sign 'plus'. Then, as an analysis of the NJL model
teaches, the branch of equation solution for negative dynamical quark mass is the most
stable one. Let us remember here we are dealing with the Euclidean metrics (though it
is not a principal point) and a quark mass appears in the corresponding expressions
as an imaginary quantity. Now substituting the calculated expressions for the
pairing angle into the trigonometrical factor
$$\sin \left(\theta-\theta_m\right)=
\sin \theta~\frac{p}{|p_4|}-\cos \theta~\frac{m}{|p_4|}~$$
and performing some algebraic transformations we come to the relation
\begin{equation}
\label{mass}
M_q({\bf p})=M({\bf p})-m~.
\end{equation}
In particular, the equation for dynamical quark mass (\ref{19}) is getting
the form characteristic for the mean field approximation
\begin{equation}
\label{23}
M=2G~\int \frac{d{\bf q}}{(2\pi)^3}
~(1-n'-\bar n')~\frac{M'_q}{|P'_4|}~I({\bf p}+{\bf q}).
\end{equation}

The second and third equations of system (\ref{18}) allow us to find the
following expressions
\begin{equation}
\label{newden}
n=\left(e^{\beta(|P_4|-\mu)}+1\right)^{-1},
~\bar n=\left(e^{\beta~(|P_4|+\mu)}+1\right)^{-1},
\end{equation}
and, hence, the thermodynamic properties of our system, in particular,
the pressure of quark ensemble
$$P=-\frac{d E}{d V}~.$$
\begin{figure}
\includegraphics[width=0.3\textwidth]{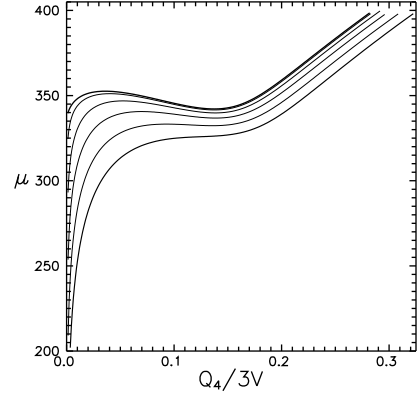}
\caption{The chemical potential $\mu$ (MeV) is plotted as a function of charge
density ${\cal Q}_4=Q_4/(3V)$ (in the units of ch/fm$^3$). The factor 3 relates the
densities of quark and baryon matters. The top curve corresponds to the situation of zero
temperature. The curves following down correspond to the temperature values
$T=10$ MeV, ... , $T=50$ MeV with spacing $T=10$ MeV.
}
\label{f2}
\end{figure}

By definition we should calculate this derivative at constant mean entropy,
$d\bar S/dV=0$. This condition makes possible, for example, to calculate the
derivative $d\mu/dV$, but the mean charge $\bar Q_4$ should not also change.
In order to maintain it valid we introduce two independent chemical
potentials --- for quarks $\mu$ and for anti-quarks $\bar\mu$ (following
Eq. (\ref{newden}) with the opposite signs).
It leads also to the change $\mu \to \bar \mu$ in definition of $\bar n$ in Eq. (ref{newden}).
This kind of description apparently allows us to treat even some non-equilibrium
states of quark ensemble (but with losing a covariance similar to the situation which
takes place in electrodynamics while one deals with electron-positron gas). Here we
are interested in the unaffected balanced situation of $\bar\mu=\mu$. Then the
corresponding derivative of specific energy $d w/d V$ might be presented as
\begin{eqnarray}
&&\frac{d w}{d V}=\int \frac{d{\bf p}}{(2\pi)^3}
\left(\frac{d n}{d\mu}\frac{d\mu}{dV}+\frac{d \bar n}{d\bar\mu}
\frac{d\bar\mu}{dV}\right)
\left[|p_4|\cos\theta-\right.\nonumber\\
&&\left.-2G \sin \left(\theta-\theta_m\right)
\int\frac{d{\bf q}}{(2\pi)^3}
\sin\left(\theta'-\theta'_m\right)(n'+\bar n'-1)I\right].\nonumber
\end{eqnarray}
Now representing the trigonometric factors via dynamical quark mass and drawing Eq.
(\ref{19}) we obtain for the ensemble pressure
\begin{equation}
\label{press}
\hspace{-0.15cm}P=-\frac{E}{V}-V2N_c\int \frac{d{\bf p}}{(2\pi)^3}
\left(\frac{d n}{d\mu}\frac{d\mu}{dV}+\frac{d \bar n}{d\bar\mu}
\frac{d\bar\mu}{dV}\right)|P_4|.
\end{equation}

\begin{figure}
\includegraphics[width=0.3\textwidth]{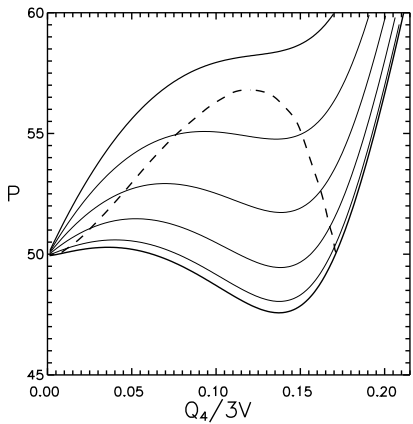}
\caption{The ensemble pressure $P$ (MeV/fm$^3$) is shown as a function of charge
density ${\cal Q}_4$ at temperatures $T=0$ MeV, ... , $T=50$ MeV with spacing
$T=10$ MeV. The lowest curve corresponds to zero temperature. The dashed curve
shows the boundary of phase transition liquid--gas, see the text.
}
\label{f2a}
\end{figure}
The requirement of mean charge conservation
\begin{equation}
\label{q4}
\hspace{-0.2cm}\frac{d \bar Q_4}{d V}=
\!\frac{\bar Q_4}{V}+V 2N_c\!\int \!\frac{d{\bf p}}{(2\pi)^3}
\left(\frac{d n}{d\mu}\frac{d\mu}{dV}-\frac{d \bar n}{d\bar\mu}
\frac{d\bar\mu}{dV}\right)=0,
\end{equation}
provides us with an equation which interrelates the derivatives $d\mu/dV$ and
$d\bar\mu/dV$. Apparently, the regularized expressions for mean charge of quarks
and anti-quarks are meant (\ref{ntot}) here. Dealing in a similar way with the
requirement of mean entropy conservation, $d \bar S/dV=0$, we receive another
equation as
\begin{eqnarray}
\label{entr}
&&\int \frac{d{\bf p}}{(2\pi)^3}\frac{d n}{d\mu}\ln \frac{n}{1-n}
\frac{d\mu}{d V} -\int \frac{d{\bf p}}{(2\pi)^3}
\frac{d \bar n}{d\bar\mu}\ln \frac{\bar n}{1-\bar n}
\frac{d\bar\mu}{d V}
=\nonumber\\
&&=\frac{\bar S}{2N_c~V^2}~.
\end{eqnarray}
Substituting here $T\ln(n^{-1}-1)=-\mu+|P_4|$ and\\
$T\ln (\bar n^{-1}-1)=\bar\mu+|P_4|$ we have after simple calculations
with taking into account (\ref{q4}) that
$$\int \frac{d{\bf p}}{(2\pi)^3}
\left(\frac{d n}{d\mu}\frac{d\mu}{dV}+\frac{d \bar n}{d\bar\mu}
\frac{d\bar\mu}{dV}\right)|P_4|=-\frac{\bar S T}{2N_c V^2}-\frac{\bar
Q_4\mu}{2N_c V^2}~.
$$
Eventually it leads to the following expression for the pressure
\begin{equation}
\label{p}
P=-\frac{E}{V}+\frac{\bar S~T }{V}+\frac{\bar Q_4~\mu}{V}~.
\end{equation}
(of course, the thermodynamic potential is $\Omega=-P~V$). At small temperatures
the anti-quark contribution is negligible, and thermodynamic description can be
grounded on utilizing one chemical potential $\mu$ only. If the anti-quark contribution
is getting intrinsic the thermodynamic picture becomes more complicated due to the
necessity to obey the condition $\bar\mu=\mu$ which comes to the play. In
particular, at zero temperature we might consider the anti-quark
contribution absent and obtain
$$P= -{\cal E}+\mu~\rho_q~,$$
where $\mu=\left(P_F^2+M^2_q(P_F)\right)^{1/2}$, $P_F$ is the Fermi momentum and
$\rho_q=N/V$ is the quark ensemble density.
\begin{figure}
\includegraphics[width=0.3\textwidth]{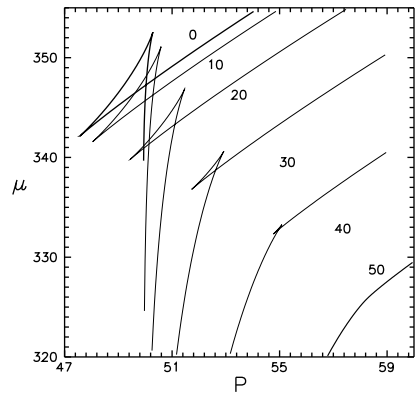}
\caption{The fragments of isotherms in Fig. \ref{f2}, \ref{f2a}, see text.
Chemical potential $\mu$ (MeV) is plotted as a function of pressure $P$ MeV/fm$^3$. The
top curve corresponds to the zero isotherm and following down with spacing 10 MeV
till the isotherm 50 MeV (the lowest curve).}
\label{f3}
\end{figure}

For lucidity of our view point we consider mainly the NJL model \cite{4} in this
paper, i.e. the correlation function (the form-factor in Eq. (\ref{cor}))
behaves as the $\delta$-function in coordinate space. It is well known fact that in
order to have an intelligent result in this model one needs to use a regularization
cutting of the momentum integration in Eq. (\ref{17}). We adjust the standard
set of parameters \cite{5} here with $|{\bf p}|<\Lambda$, $\Lambda=631$ MeV,
$m=5.5$ MeV
and $G\Lambda^2/(2\pi^2)=1.3$. This set at $n=0$, $\bar n=0$, $T=0$ gives for
the dynamical quark mass $M_q=335$ MeV. In particular, it may be shown the following
representation of ensemble energy is valid at the extremals of functional (\ref{17})
\begin{eqnarray}
\label{mean}
&&\hspace{-0.3cm}E=E_{vac}+2 N_cV\int^\Lambda\frac{d{\bf p}}{(2\pi)^3}~|P_4|~(n+\bar
n)~,\nonumber\\
[-.2cm]
\\ [-.25cm]
&&\hspace{-0.3cm}E_{vac}=2 N_c V\!\!\! \int^\Lambda \frac{d{\bf p}}{(2\pi)^3}
(|p_4|- |P_4|)+2 N_c V \frac{M^2}{4 G},\nonumber
\end{eqnarray}
It is easy to understand this expression with the vacuum contribution subtracted
looks like the energy of a gas of relativistic particles and antiparticles with the
mass $M_q$ and coincides identically with that calculated in the mean field
approximation.
\begin{figure}
\includegraphics[width=0.3\textwidth]{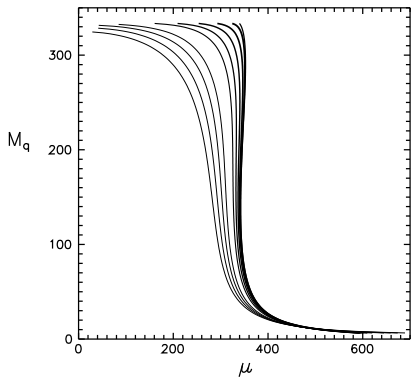}
\caption{The dynamical quark mass $|M_q|$ (MeV) is shown as a function of
chemical potential $\mu$ (MeV) at the temperatures $T=0$ MeV, ... , $T=100$ MeV with spacing
$T=10$ MeV. The most right curve corresponds to zero temperature.
}
\label{f4}
\end{figure}

Let us summarize the results of this exercise. So, we determine the density of
quark $n$ and anti-quark $\bar n$ quasi-particles at given parameters $\mu$ and $T$ from the
second and third equations of system (\ref{18}). From the first equation we receive the
angle of quark and anti-quark pairing $\theta$ as a function of dynamical quark mass
$M_q$ which is handled as a parameter. Then at small temperatures, below $50$ MeV, and value of
chemical potentials of dynamical quark mass order, $\mu\sim M_q$, there are several branches of
solutions of the gap equation. Fig. \ref{f1} displays the difference of right and left sides of
Eq. (\ref{19}) which is denoted by $\Delta$ at zero temperature and several values of chemical
potential $\mu$ (MeV) = $335$ (the lowest curve), $340$, $350$, $360$ (the top curve) as a
function of parameter $M_q$. The zeros of function $\Delta(M_q)$ correspond to the
equilibrium values of dynamical quark mass.

The evolution of chemical potential as a function of charge density
${\cal Q}_4=Q_4/(3V)$ (in the units of charge/$fm^3$) with the temperature
increasing is depicted in Fig. \ref{f2}
(factor 3 relates the quark and baryon matter densities). The top curve
corresponds to the zero temperature. The other curves following down have been calculated for the
temperatures $T=10$ MeV, ... , $T=50$ MeV with spacing $T=10$ MeV. As it was found in
Ref. \cite{ff} the chemical potential at zero temperature is increasing first with the charge
density increasing, reaches its maximal value, then decreases and at the densities of order of
normal nuclear matter density\footnote{At the Fermi momenta of dynamical quark mass order.},
$\rho_q\sim 0.16/fm^3$, becomes almost equal its vacuum value. Such a behaviour
of chemical potential results from the fast decrease of dynamical quark mass with the Fermi
momentum increasing. It is clear from Fig. \ref{f2} the charge density is still
multivalued function of chemical potential at the temperature slightly below $50$ MeV.
The Fig. \ref{f2a} shows the ensemble pressure $P$ (MeV/fm$^3$) as the function of charge
density ${\cal Q}_4$ at several values of temperature. The lowest curve corresponds to the zero
temperature. The other curves following up correspond to the temperatures $T=10$ MeV, ... ,
$T=50$ MeV (the top curve) with spacing $T=10$ MeV. It is ~curious to remember now that in
Ref. \cite{ff} the vacuum pressure estimate for the NJL model was received as
$40$---$50$ MeV/fm$^3$ which is entirely compatible with the results of conventional
bag model. Besides, some hints at an instability presence (rooted in the anomalous
behavior of pressure $dP/dn<0$, see also
\cite{T}, \cite{M}) in some interval of the Fermi momentum has been found out.

Fig. \ref{f3} shows the fragments of isotherms of Fig. \ref{f2}, \ref{f2a} but
in the different coordinates (chemical potential --- ensemble pressure). The top curve is
calculated at the zero temperature, the other isotherms following
down correspond to the temperatures increasing with
spacing 10 MeV. The lowest curve is calculated at the temperature 50 MeV. This
plot obviously demonstrates that there are the states on the isotherm which are
thermodynamically equilibrated and have an equal pressure and chemical potential
(see the characteristic Van der Waals triangle with the crossing curves).
The equilibrium points calculated are shown in Fig. \ref{f2a} by the
dashed curve. The points of dashed curve crossing with an isotherm pin-point the
boundary of gas---liquid phase transition. The corresponding straight line
$P=\mbox{const}$ which obeys the Maxwell rule separates the
non-equilibrium and unstable fragments of isotherm and describes a
mixed phase. Then the critical temperature for the parameter which we are using
in this paper becomes $T_{c}\sim 45.7$ MeV with the critical charge density as $\bar Q_4\sim
0.12$ ch/fm$^3$. Usually the thermodynamic description is grounded on the mean energy functional
which is the homogeneous function of particle number like $E=N~f(S/N,V/N)$ (without vacuum
contribution). It is clear such a description requires the corresponding subtractions to be
introduced, however, this operation does not change the final results considerably. Now the vague
arguments of Refs. \cite{ff} that the states filled up with quarks and separated from the
instability region look like a 'natural construction material' to form the baryons
are getting much more clarity and give a hope to understand the existing fact
of equilibrium between vacuum and octet of stable (in strong interaction)
baryons\footnote{The chiral quark condensate for the filled up
state develops the quantity about (100 MeV)$^3$ (at $T=0$), see \cite{ff}, that demonstrates the
obvious tendency of restoring a chiral symmetry.}.

\begin{figure}
\includegraphics[width=0.3\textwidth]{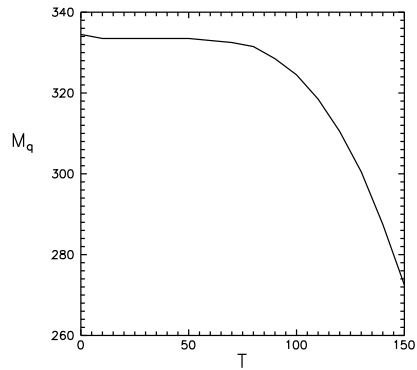}
\caption{ The dynamical quark mass $|M_q|$ (MeV) as a function of temperature at
the small value of charge density ${\cal Q}_4$.}
\label{f4a}
\end{figure}

The dynamical quark mass $|M_q|$ (MeV) as a function of chemical potential
$\mu$ (MeV) is presented for the temperatures $T=0$ MeV, ... , $T=100$ MeV with spacing
$T=10$ MeV in Fig. \ref{f4}. The most right-hand curve corresponds to the zero temperature.
At small temperatures, below $50$ MeV, the dynamical quark mass is the multivalued
function of chemical potential. Fig. \ref{f4a} shows
the dynamical quark mass as a function of temperature at small values of charge
density ${\cal Q}_4\sim 0$. Such a behaviour allows us to conclude that the
quasi-particle size is getting larger with temperature increasing.
It becomes clear if we remember that the momentum corresponding
the maximal attraction between quark and anti-quark $p_\theta$ (according to
Ref. \cite{MZ}) is defined by $d\sin \theta/d p=0$. In particular, this parameter
in the NJL model equals to
\begin{equation}
\label{pt}
p_\theta=(|M_q|~m)^{1/2}~.
\end{equation}
but its inverse magnitude defines the characteristic (effective) size of quasi-particle
$r_\theta=p_\theta^{-1}$.

If one is going to define the quark chemical potential as an energy necessary to
add (to remove) one quasi-particle (as it was shown in \cite{ff} at zero temperature),
$\mu=dE/dN$, then in vacuum (i.e. at quark density $\rho_q$ going to zero)
quark chemical potential magnitude coincides with the
quark dynamical mass. It results in the phase diagram displayed at this value of
chemical potential although, in principle, this value could be smaller
than dynamical quark mass as it has been considered in the pioneering paper \cite{ayz}.
If one takes, for example, chemical potential value equal to zero it leads to
the conventional picture but, obviously, such a configuration does not
correspond to the real process of filling up the Fermi sphere with quarks.

Apparently, our study of the quark ensemble thermodynamics produces quite
reasonable arguments to propound the hypothesis that the phase transition of chiral symmetry
(partial) restoration has been already realized as the mixed phase of physical
vacuum and baryonic matter{\footnote{Indirect confirmation of this hypothesis one
could observe, for example, in the existing degeneracy of excited
baryon states Ref. \cite{Gloz}.}}. However, it is clear our quantitative
estimates should not be taken as appropriate for comparing with, for example,
the critical temperature of nuclear matter
phase transition which has been experimentally measured and is equal $15$--$20$ MeV.
Besides, the gas component (at $T=0$) has nonzero density ($0.01$ of the normal nuclear density)
but in reality this branch should correspond to the physical vacuum, i.e. zero baryonic
density{\footnote{Similar uncertainty is present in the other predictions of chiral
symmetry restoration scenarios, for example, it stretches from $2$ till $6$ units
of the normal nuclear density.}}. In principle, an idea of global
equilibrium of gas and liquid phases pursued us to put the adequate boundary
conditions down at describing the transitional layer existing between the vacuum and the filled-up
state and to calculate the surface tension effects. It looks plausible that the changes taking
place in this layer could ascertain all ensemble processes similar to the theory of Fermi-liquids.

\section{Transition layer between gas and liquid}
This concept advanced would obtain the substantial confirmation if we are able
to demonstrate an existence of this transition layer at which the ensemble
transformation from one aggregate state to
another takes place. As it was argued above the indicative characteristic to
explore a homogeneous phase (at finite temperature) is the mean charge (density)
of ensemble. It was demonstrated the other characteristics, for example, a chiral condensate, a
dynamical quark mass, etc. could be reconstructed as well.
 So, here we are analyzing the transition layer at zero temperature.

If one assumes the parameters of gas phase are approximately the same as those at zero
charge $\rho_g=0$, i.e. as in vacuum
(it means ignoring the negligible distinctions in the pressure, chemical
potential and quark condensate),  the dynamical quark mass develops the
maximal value, and for the parameter choice of the NJL model it is $M=335$ MeV.
Then from the Van der Waals diagram one may
conclude that the liquid phase being in equilibrium with the gas phase develops
the density $\rho_l=3\times 0.185$ ch/fm$^3$ (by some reason which becomes clear below we
correct it in favour of the value $\rho_l=3\times 0.157$ ch/fm$^3$). The detached factor $3$
here links again the magnitudes of quark and baryon densities. The quark mass is approximately
$\stackrel{*}{M}\approx 70$ MeV in this phase (and we are dealing with the simple
one-dimensional picture hereafter).

The precursor experience teaches that an adequate description of heterogeneous
states can be reached with the mean field approximation \cite{Lar}.
In our particular case it means making use the
corresponding effective quark-meson Lagrangian \cite{GL} (the functional of
Ginzburg-Landau type)
\begin{eqnarray}
\label{mesons}
&&{\cal L}=-\bar q~(\hat \partial +M)~q- \frac12~(\partial_\mu \sigma)^2-
U(\sigma)-\nonumber\\[-.2cm]
\\ [-.25cm]
&&-\frac14~F_{\mu\nu}F_{\mu\nu}-\frac{m_v^2}{2}~V_\mu V_\mu-
g_\sigma~\bar q q~\sigma+i g_v~\bar q~\gamma_\mu~q~ V_\mu~,\nonumber
\end{eqnarray}
where
$$F_{\mu\nu}=\partial_{\mu} V_\nu-\partial_{\nu} V_\mu~,~~~
U(\sigma)=\frac{m_\sigma^2}{2} ~\sigma^2+ \frac{b}{3}~\sigma^3
+\frac{c}{4}~\sigma^4~,$$
and $\sigma$ is the scalar field, $V_\mu$ is the field of vector mesons,
$m_\sigma$, $m_v$ are the masses of scalar and vector mesons and
$g_\sigma$, $g_v$ are the coupling constants of quark-meson
interaction. The $U(\sigma)$ potential includes the nonlinear terms of sigma field interactions
up to the fourth order. For the sake of simplicity we do not include the
contribution coming from the pseudoscalar and axial-vector mesons.

The meson component of such a Lagrangian should be selfconsistently treated by
considering the corresponding quark loops. (In terms of a relativistic extension of the Landau
theory of Fermi-liquid the density fluctuations (meson field collective modes) are nothing
more than the zero-sound as was shown in Ref. \cite{M}). Here we do not see any reason to go
beyond the well elaborated and reliable one loop approximation (\ref{mesons}) \cite{GL},
although recently the considerable progress has been reached (as we mention at
the beginning of this paper) in scrutinizing the nonhomogeneous quark condensates
by application of the powerful methods of exact
integration \cite{KN}. Here we believe it is more practical to adjust
phenomenologically the effective Lagrangian parameters basing on the transparent
physical picture. It is easy to see that handling (\ref{mesons}) in one loop
approximation we come, in actual fact, to the Walecka model
\cite{wal} but adopted for the quarks. In what follows we are working with the
designations of that model and do hope it does not lead to the misunderstandings.

In the context of our paper we propose to interpret Eq. (\ref{mesons}) in the
following way. Each phase might be considered, in a sense, with regard to another phase as an
excited state which requires the additional (apart from a charge density) set of parameters
(for example, the meson fields) for its complete description, and those are
characterizing the measure of deviation from the equilibrium state.
Then the crucial question becomes whether it is possible to adjust the
parameters of effective Lagrangian (\ref{mesons}) to obtain the solutions
in which the quark field interpolates between the
quasi-particles in the gas (vacuum) phase and the quasi-particles of the filled-up states.
For all that the density of the filled-up state ensemble should asymptotically approach the
equilibrium value of $\rho_l$ and should turn to the zero value in the gas phase (vacuum).

The scale inherent in this problem may be assigned with one of the mass referred
in the Lagrangian (\ref{mesons}). In particular, we bear in mind the dynamical quark mass in the
vacuum $M$. Besides, there are another four independent parameters in the problem and in order to
compare them with the results of studying a nuclear matter we employ the form
characteristic for the (nuclear) Walecka model
$$C_s=g_\sigma~\frac{M}{m_\sigma}~,~~C_v=g_v~\frac{M}{m_v}~,
~~\bar b=\frac{b}{g_\sigma^3~M}~,~~\bar c=\frac{c}{g_\sigma^4}~.$$
Taking the parameterization of the potential $U(\sigma)$ as
$b_\sigma=1.5~m_\sigma^2~(g_\sigma/M)$,
$c_\sigma=0.5~m_\sigma^2~(g_\sigma/M)^2$ we come to the sigma model but the
choice $b=0$, $c=0$ results in the Walecka model. As to standard nuclear matter application the
parameters $b$ and $c$ demonstrate vital model dependent character and are quite different from the
parameter values of sigma model. ~Truly, in that case their values are also regulated by additional
requirement of an accurate description of the saturation property. On the other hand, for the quark
Lagrangian (\ref{mesons}) we could intuitively anticipate some resemblance with the sigma model and, hence,
could introduce two dimensionless parameters $\eta$ and $\zeta$ in the form of $b=\eta~b_\sigma$,
$c=\zeta^2~c_\sigma$ which characterize some fluctuations of the effective potential.
Then the scalar field potential is presented as follows
$$U(\sigma)=\frac{m_\sigma^2}{8}~\frac{g_\sigma^2}{M^2}~
\left(4~\frac{M^2}{g_\sigma^2}+4~\frac{M}{g_\sigma}~
\eta~\sigma+\zeta^2\sigma^2\right)~\sigma^2~.$$

The meson and quark fields are defined by the following system of the stationary equations
\begin{eqnarray}
\label{sys}
&&\Delta~ \sigma -
m_\sigma^2~\sigma=b~\sigma^2+c~\sigma^3+g_\sigma~\rho_s~,\nonumber\\
&&\Delta ~V - m_v^2~ V=-g_v~\rho~,\\
&&( {\bf \hat\nabla}+\stackrel{*}{M})~q=(E-g_v~V)~q~,\nonumber
\end{eqnarray}
where $\stackrel{*}{M}=M+g_\sigma\sigma$ is the running value of dynamical quark
mass, $E$ stands for the quark energy and $V=-iV_4$. The density matrix describing the quark
ensemble at $T=0$ has the form
$$ \xi (x)=\int^{P_F}\frac{{d\bf p}}{(2\pi)^3}~q_{\bf p}(x)~
\bar q_{\bf p}(x)~,$$
in which ${\bf p}$ is the quasi-particle momentum and the Fermi momentum
$P_F$ is defined by the corresponding chemical potential.
The densities $\rho_s$ and $\rho$ at the right hand sides of
Eq. (\ref{sys}) are by definition
$$\rho_s(x)=Tr\left\{\xi(x),1
\right\}~,~~\rho(x)=Tr\left\{\xi(x),\gamma_4\right\}~.$$

Here we confine ourselves to the Thomas--Fermi approximation while describing
the quark ensemble. Then the densities which we are interested in are
given with some local Fermi momentum $P_F(x)$ as
\begin{eqnarray}
\label{rhorhos}
&&\hspace{-0.5cm}\rho= \gamma\int^{P_F}\frac{d {\bf p}}{(2\pi)^3}=
\frac{\gamma}{6\pi^2}~P_F^3~,\\
&&\hspace{-0.5cm}\rho_s=\gamma\int^{P_F}\frac{d {\bf p}}{(2\pi)^3}~
\frac{\stackrel{*}{M}}{E}=\nonumber\\
&&\hspace{-0.5cm}=\frac{\gamma}{4\pi^2}\stackrel{*}{M}P_F^2
\left\{\left(1+\lambda^2\right)^{1/2}-\frac{\lambda^2}{2}
\ln\left[\frac{\left(1+\lambda^2\right)^{1/2}+1}
{\left(1+\lambda^2\right)^{1/2}-1}\right]\right\},\nonumber
\end{eqnarray}
where $\gamma$ is the quark factor which for one flavour is $\gamma=2
N_c$ ($N_c$ is the number of colours), $E=({\bf p}^2+\stackrel{*}{M}^2)^{1/2}$ and
$\lambda=\stackrel{*}{M}/P_F$. By definition the ensemble chemical potential
does not change and it leads to the situation in which the local value
of Fermi momentum is defined by the running value of dynamical quark mass and vector field as
\begin{equation}
\label{mu}
\mu=M=g_v~V+\left(P_F^2+\stackrel{*}{M}^2\right)^{1/2}~.
\end{equation}

Now we should tune the Lagrangian parameters (\ref{mesons}). For asymptotically
large distances (in the homogeneous phase) we may neglect the gradients of scalar and vector
fields and the equation for scalar field in the system (\ref{sys}) leads to the first
equation which bounds the parameters $C_s$, $C_v$, $\bar b$, $\bar c$
\begin{equation}
\label{id1}
\hspace{-0.1cm}
\frat{\stackrel{}{M}^2 \!\!\!\!(\stackrel{*}{M}-
\stackrel{}{M})}{C_s^2}+\bar b\stackrel{}{M}\!\!\!(\stackrel{*}{M}-
\stackrel{}{M})^2
+\bar c(\stackrel{*}{M}-
\stackrel{}{M})^3=-\rho_s.
\end{equation}
The asymptotic vector field is given by the ensemble density
$V=C_v^2~\rho/(g_v M^2)$. The second equation results from the
relation (\ref{mu}) for the chemical potential and gives
\begin{equation}
\label{id2}
M=\frac{C_v^2~\rho}{M^2}+\left(P_F^2+\stackrel{*}{M}^2\right)^{1/2}~.
\end{equation}

Extracting the liquid density from (\ref{rhorhos}) we obtain the Fermi
momentum ($P_F=346$ MeV) and applying the identities (\ref{id1}), (\ref{id2})
we have for the particular case $b=0$, $c=0$ that $C_s^2=25.3$, $C_v^2=-0.471$,
i.e. the vector component $C_v^2$ is small (comparing to $C_s^2$)
and has negative value which is unacceptable. Apparently, it looks necessary to
neglect the contribution coming from the vector field or to diminish the dynamical quark
mass $\stackrel{*}{M}$ up to the value which retains the identity (\ref{id2}) valid
with positive $C_v^2$ or equals to zero. In the gas phase the dynamical quark mass
can also be corrected to the value larger than the vacuum value.
It is clear that in the situation of the liquid with the density $\rho_l=3\times 0.185$ ch/fm$^3$
the dynamical quark mass should coincide (or exceed) $M=346$ MeV in the gas
phase. However, here we correct the liquid density (as it was argued above)
to decrease its value up to $\rho_l=3\times 0.157$
ch/fm$^3$ which is quite acceptable in the capacity of normal nuclear matter
density. In fact, this possibility can be simply justified by another choice of the NJL model
parameters. Thus, we obtain at
$\stackrel{*}{M}=70$ МэВ and $b=0$, $c=0$ that $C_s^2=28.4$, $C_v^2=0.015$, i.e.
we have a small but positive value for the vector field coefficient. At the same time,
being targeted here to estimate the surface
tension effects only we do not strive for the precise fit of parameters. In the
Walecka model these coefficients are $C_s^2=266.9$, $C_v^2=145.7$, ($b=0$, $c=0$).
Moreover, there is another parameter set
with $C_s^2=64.$, $C_v^2\approx 0$ \cite{Bog} but it is rooted in an essential
nonlinearity of the sigma-field due to the nontrivial values of the coefficients $b$ and $c$.
The option (formally unstable) with negative $c$ ($b$) has been also discussed.

\begin{figure}
\begin{center}
\leavevmode
\hspace{-1.5cm}
\includegraphics[width=0.3\textwidth]{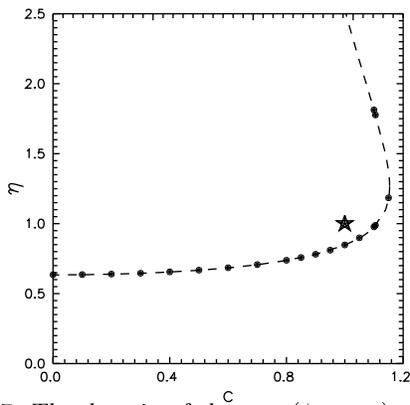}
\vspace{-5mm}
\caption{The domain of the $\eta$, $c$ ($\zeta=c~\eta$)-plane in which an
increase of specific energy occurs, see the text. The dots represent a stable kink.
The star shows the position of canonical (chiral) kink, see the text.
}
\label{f5}
\end{center}
\end{figure}

The coupling constant of scalar field is fixed by the standard (for the NJL
model) relation between the quark mass and the $\pi$-meson decay constant
$g_\sigma=M/f_\pi$ (we put $f_\pi=100$ МeV) although there
is no any objection to treat this coupling constant as an independent parameter.
As a result of all agreements done we have for the $\sigma$-meson mass
$m_\sigma=g_\sigma~M/C_s$. In principle, we could even
fix the $\sigma$-meson mass and coupling constant $g_\sigma$ but all relations
above mentioned lead eventually to quite suitable values of the $\sigma$-meson mass as will be
demonstrated below.

The vector field plays, as we see, a secondary role because of the small
magnitude of constant $C_v$. Then taking the vector meson mass as
$m_v\approx 740$ MeV (slightly smaller than the mass of $\omega$-meson
because of simple technical reason only) we calculate the coupling constant of
vector field from the relation similar to the scalar field $m_v=g_v~M/C_v$.
Amazingly, its value comes about steadily small in
comparing to the value characteristic for the NJL model $g_v=\sqrt{6} g_\sigma$.
However, at the values of constant $C_v$ which we are interested in it is very
difficult to maintain the reasonable balance and for
the sake of clarity we prefer to choose the massive vector field. Actually, it
is unessential because we need this parameter (as we remember) only to estimate
the vector field strength.

\begin{figure}
\begin{center}
\leavevmode
\includegraphics[width=0.3\textwidth]{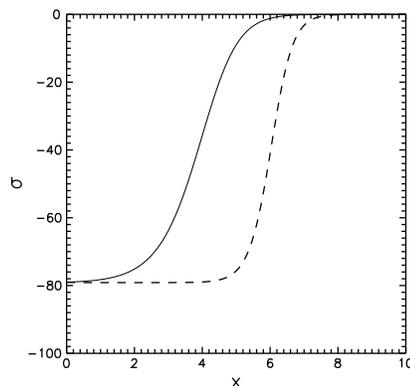}
\vspace{-5mm}
\caption{The stable kink solutions with $c=1.1$, the solid line
corresponds to $\eta\approx 0.977$ ($m_\sigma\approx 468$ MeV) and the dashed
line corresponds to $\eta\approx 1.813$ ($m_\sigma\approx 690$ MeV),
$x$ is given in the units of $fm$ and $\sigma$ is given in MeV.
}
\label{f5a}
\end{center}
\end{figure}
The key point of our interest here is the surface tension coefficient \cite{Bog}
which can be defined as
\begin{equation}
\label{sur}
u_s=4\pi~r_o^2~\int_{-\infty}^{\infty}dx~
\left[ {\cal E}(x)-\frac{{\cal E}_l}{\rho_l}~\rho(x)\right]~.
\end{equation}
The parameter $r_o$ will be discussed in the next section at considering the
features of quark liquid droplet, and for the present we would like to notice only that for the
parameters considered its magnitude for $N_f=1$ is around $r_o=0.79$ fm.
Recalling the factor $3^{1/3}$ which connects the baryon and quark
numbers one can find the magnitude ($\widetilde r_o=3^{1/3} 0.79\approx
1.14$ fm) is in full agreement with the magnitude standard for the
nuclear matter calculations (in the Walecka model)
$\widetilde r_o=1.1$ --- $1.3$ fm.

In order to proceed we calculate ${\cal E}(x)$ in the Thomas--Fermi approximation as
\begin{eqnarray}
{\cal E}(x)&=&\gamma~\int^{P_F(x)}\frac{d {\bf p}}{(2\pi)^3}~
[{\bf p}^2+\stackrel{*}{M}(x)]^{1/2}+\nonumber\\
&+&\frac12~g_v~\rho(x)~V(x)-
\frac12~g_\sigma~\rho_s(x)~\sigma(x)~.\nonumber
\end{eqnarray}
And to give some idea for the 'setup' prepared we present here the
characteristic parameter values for
some fixed $b$ and $c$ with $\rho_l=3\times 0.157$ ch/fm$^3$. In the liquid
phase they are
$\stackrel{*}{M}=70$ MeV ($P_F=327$ MeV) and $e_l=310.5$ MeV (index $l$ stands
for a liquid phase and $e(x)={\cal E}(x)/\rho(x)$ defines the density of specific energy).
Both equations (\ref{id1}) and (\ref{id2}) are obeyed by this state. There exist
the solution with larger value of quark mass $\stackrel{*}{M}=306$ MeV, ($P_F=135$ MeV) (we have
faced the similar situation in the first section dealing with the gas of quark quasi-particles)
and $e=338$ MeV $\sim e_g$ ($e_g$ is the specific energy in the gas phase)
which satisfies both equations as well. The specific energy of this
solution occurs to be larger than specific energy of the previous solution.
It is worthwhile to mention the existence of intermediate state corresponding
to the saturation point with the mass
$\stackrel{*}{M}=95$ MeV, ($P_F=291$ MeV) and $e=306$ MeV. Obviously, it is
the most favorable state with the smallest value of specific energy
(and with the zero pressure of quark ensemble), and the system
can reach this state only in the presence of significant vector field.
This state (already discussed in the first section) corresponds to the minimal
value of chemical potential ($T=0$) and can be reached at
the densities typical for the normal nuclear matter. However, Eq. (\ref{id2}) is
not valid for this state.

Two another parameters $\eta$, $\zeta$ are fixed by looking through all the
configurations in which the solution of equation system (\ref{sys})
with stable kink of the scalar field does exist and describes
the transition of the gas phase quarks to the liquid phase. First, it is
reasonable to scan the $\eta$, $c$ ($\zeta=c~\eta$)-plane, in order
to identify the domain in which the increase of specific energy
${\cal E}-{\cal E}_l~\rho/\rho_l\le 0$ is revealed at running through all
possible states which provide the necessary transition (without taking
into account the field gradients). In practice one need to follow
a simple heuristic rule. The state with $P_F \sim 1$ MeV (i.e. $e$ and the
corresponding $\rho$). The state of characteristic liquid energy
${\cal E}_l$ (together with $\rho_l$) should be compared at scanning
the Lagrangian parameters $\eta$ and $c$. Just this domain where they are
commensurable could provide us with the solutions in which we are interested
and Fig.\ref{f5} shows its boundary. The curve could be continued
beyond the value $\eta=2.5$ but the values of corresponding parameter $\eta$ are
unrealistic and not shown in the plot.

We calculate the solution of equation system (\ref{sys}) numerically by the
Runge--Kutta method with the initial conditions $\sigma(L)\approx 0$,
$\sigma'(L)\approx 0$ imposed at the large distance $L\gg t$
where $t$ is a characteristic thickness of transition layer (about 2 fm). Such a
simple algorithm occurs quite suitable if the vector field contribution
is considered as a small correction (what just takes place
in the situation under consideration) and is presented as
$$V(x)=\frac{1}{2m_v}~\int^L_{-L}~dz~e^{-m_v|x-z|}~g_v~\rho(z)~,$$
where the charge (density) $\rho$ is directly defined by the scalar field. We
considered the solutions including the contribution of the vector field as well
and the corresponding results confirm the estimates obtained.

\begin{figure}[!tbh]
\begin{center}
\leavevmode
\includegraphics[width=0.3\textwidth]{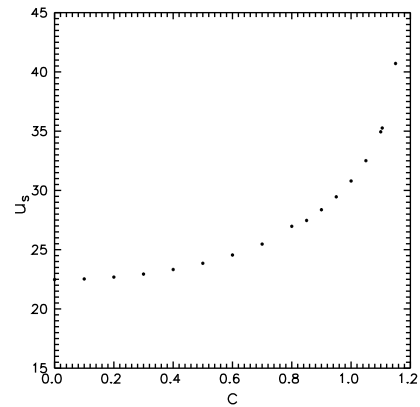}
\caption{The surface tension coefficient $u_s$ in MeV as a function
of parameter $c$ ($\zeta=c~\eta$) for the curve of stable kinks (with $\eta\leq 1.2$).}
\label{f6}
\end{center}
\end{figure}

Rather simple analysis shows the interesting solutions are located along the
boundary of discussed domain. Some of those are depicted in Fig. \ref{f5}
as the dots. Fig. \ref{f5a} shows the stable kinks of
$\sigma$-field with the parameter $c=1.1$ for two existing solutions with
$\eta\approx 0.977$ ($m_\sigma\approx 468$ MeV) (solid line) and $\eta\approx
1.813$ ($m_\sigma\approx 690$ MeV) (dashed line). For the sake of clarity
we consider the gas (vacuum) phase is on the right. Then the asymptotic value of
$\sigma$-field on the left hand side ($\sigma\approx 80$ MeV) corresponds to
$\stackrel{*}{M}=70$ MeV. The thickness of transition layer for the solution
with $\eta\approx 0.977$ is $t\approx 2$ fm while for the second solution  $t\approx 1$ fm.

Characterizing the whole spectrum of the solutions obtained we should mention
that there exist another more rigid (chiral) kinks which correspond
to the transition into the state with the dynamical quark mass
changing its sign, i.e. $M\to-M$. In particular, the kink with the canonical
parameter values $\eta=1$, $c=1$ is clearly seen (marked by the star in
Fig. \ref{f5}) and its surface tension coefficient is about
$2m_\pi$ ($m_\pi$ is the $\pi$-meson mass). The most populated class of
solutions consists of those having the meta-stable character.
The system comes back to the starting point (after an evolution) pretty rapidly,
and usually the $\sigma$-field does not evolve in such an extent to reach the
asymptotic value (which corresponds to the dynamical quark mass in the liquid phase
$\stackrel{*}{M}=70$ MeV). Switching on the vector field changes the solutions
insignificantly (for our situation with small $C_v$ it does not exceed $2$ MeV in the maximum).

\begin{figure}
\begin{center}
\leavevmode
\hspace{-1.5cm}
\includegraphics[width=0.3\textwidth]{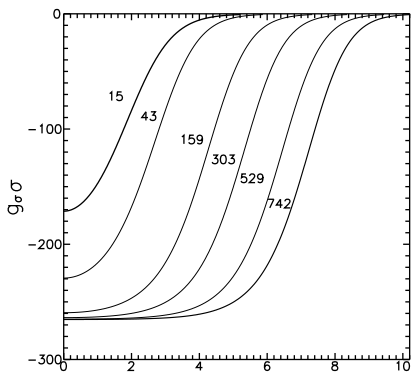}
\vspace{-5mm}
\caption{$\sigma$-field (MeV) as a function of the distance $r$ (fm) for several
solutions of the equation system (\ref{sys}) which are characterized by the net
quark number $N_q$ written to the left of each curve.
}
\label{f7}
\end{center}
\end{figure}
The surface tension coefficient $u_s$ in MeV for the curve of stable kinks with
parameter $\eta\leq 1.2$ as the function of parameter $c$ ($\zeta=c~\eta$)
is depicted in Fig. \ref{f6}. The $\sigma$-meson mass at
$c\approx 0$ is $m_\sigma\approx 420$ MeV and changes smoothly up to the value
$m_\sigma\approx 500$ MeV at $c\approx 1.16$ (the maximal value of
the coefficient $c$ beyond which the stable kink solutions are not
observed). In particular, $m_\sigma\approx 450$ MeV at $c=1$.
Two kink solutions with $c=1.1$ for $\eta \approx 0.977$ and for
$\eta\approx 1.813$ (shown in Fig. \ref{f5a}, and the second one is not shown in
Fig. \ref{f6}) have the tension coefficient values $u_s \approx 35$ MeV and $u_s
\approx 65$ MeV, correspondingly.
The maximal value of tension coefficient
for the normal nuclear matter does not exceed $u_s=50$ MeV. The nuclear Walecka model
claims the value $u_s\approx 19$ MeV \cite{Bog} as acceptable and
calculable. The reason to have this higher value of surface tension coefficient
for quarks is rooted in the different magnitudes of the mass deficit.
Indeed, for nuclear matter it does not exceed
$\stackrel{*}{M}\approx 0.5 M$ albeit more realistic values are considered
around $\stackrel{*}{M}\approx 0.7 M$ and for the quark ensemble the mass deficit
amounts to $\stackrel{*}{M}\approx 0.3 M$. We are also able to estimate the compression
coefficient of quark matter $K$ which occurs significantly larger than the nuclear one.
Actually, we see quite smooth analogy between the results of Section II and the results
of bag soliton model \cite{sbm}. The thermodynamic treatment developed
in the present paper allows us to formulate the adequate boundary conditions for the
bag in physical vacuum and to diminish considerably the uncertainties in searching
the true soliton Lagrangian. We believe it was also
shown here that to single out one soliton solution among others (including even
those obtained by the exact integration method \cite{KN}), which describes the transitional layer
between two media, is not easy problem if the boundary conditions above formulated
are not properly imposed.

\section{Droplet of quark liquid}
The results of two previous sections have led us to put the
challenging question about the creation and properties of finite quark
systems or the droplets of quark liquid which are in equilibrium
with the vacuum state. Thus, as a droplet we imply the spherically-symmetric
solution of the equation system (\ref{sys}) for $\sigma(r)$ and $V(r)$ with the obvious boundary
conditions $\sigma'(0)=0$ and $V'(0)=0$ in the origin (the primed variables denote the
first derivatives in $r$) and rapidly decreasing
at the large distances $\sigma \to 0$, $V \to 0$, when $r\to \infty$.

A quantitative analysis of similar nuclear physics models which includes the
detailed tuning of parameters is usually based on the comprehensive fitting
of available experimental data. This way is obviously
irrelevant in studying the quark liquid droplets. This global difficulty
dictates a specific tactics of analyzing. We propose to start, first of all,
with selecting the parameters which could be worthwhile to play a role of physical observables.
Naturally, the total baryon number which phenomenologically (via factor $3$) related
to the number of valence quark in an ensemble is a reasonable candidate for this role.
Besides, the density of quark ensemble $\rho(r)$, the mean size of droplet
$R_0$ and the thickness of surface layer $t$ look suitable for such an analysis.

It is argued above that the vector field contribution is negligible because of
the smallness of the coefficient $C_v$ comparing to the $C_s$ magnitude, and we follow this
conclusion (or assumption) albeit understand it is scarcely justified in the context
of finite quark system. Thus, we will put $g_v=0$, $V=0$ in what follows and
it will simplify all the calculations enormously.

\begin{figure}
\begin{center}
\leavevmode
\includegraphics[width=0.3\textwidth]{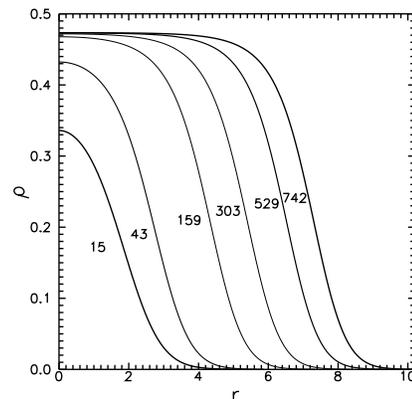}
\vspace{-5mm}
\caption{
Distribution of the quark density $\rho$ (ch/fm$^3$) for the corresponding
solutions presented in Fig. \ref{f7a}.
}
\label{f7a}
\end{center}
\end{figure}
Fig. \ref{f7} shows the set of solutions ($\sigma$-field in MeV) of the system
(\ref{sys}) at $N_f=1$ and Fig. \ref{f7a} presents the corresponding
distributions of ensemble density $\rho$ (ch/fm$^3$). The
parameters $C_s$, $C_v$, $b$ and $c$ are derived by the same algorithm as in the
previous section, i.e. the chemical potential of quark ensemble $M=335$ МэВ
(and $\sigma \to 0$) is fixed at the spatial infinity. The
filled-up states (liquid) are characterized by the parameters
$\stackrel{*}M=70$ MeV, $\rho_0=\rho_l=3\times 0.157$ ch/fm$^3$. The $\sigma$-meson mass and the
coupling constant $g_\sigma$ are derived at fixed coefficients $\eta$ and $\zeta$,
and they just define the behaviour of solutions $\sigma(r)$,
$\rho(r)$, etc. The magnitudes of functions $\sigma(r)$ and $\rho(r)$ at origin
are not strongly correlated with the values characteristic for the filled-up
states and are practically determined by solving the boundary
value problem for system (\ref{sys}). In particular, the solutions presented in
Fig. \ref{f7} have been received with the running coefficient $\eta$ at $\zeta=\eta$.
The most relevant parameter (instead of $\eta$) from the physical view point
is the total number of quarks in the droplet $N_q$ (as discussed above) and it is
depicted to the left of each curve. (The variation of $\stackrel{*}M$,
$\rho_0$ and $f_\pi$ could be considered as well instead of two mentioned
parameters $\eta$ and $\zeta$.)

Analyzing the full spectrum of solutions obtained by scanning one can reveal a
recurrent picture (at a certain scale) of kink-droplets which are easily
parameterized by the total number of quarks $N_q$ in a droplet and
by the density $\rho_0$. These characteristics are evidently fixed at completing
the calculations. The sign which allows us to single out these solutions
is related to the value of droplet specific energy (see below).

Table \ref{tab:table1} exhibits the results of fitting the density $\rho(r)$ with the Fermi
distribution
\begin{equation}
\label{ferdis}
\rho_F(r)=\frac{\widetilde \rho_0}{1+e^{(R_0-r)/b}}~,
\end{equation}
where $\widetilde \rho_0$ is the density in origin, $R_0$ is the mean size of
droplet and the parameter $b$ defines the thickness of surface layer $t=4\ln(3) b$.
Besides, the coefficient $r_0$ which is absorbed in the surface tension
coefficient (\ref{sur}), the $\sigma$-meson mass, $R_0=r_0 N_q^{1/3}$ and the coefficient
$\eta$ at which all other values have been obtained are also presented
in the Table \ref{tab:table1}.
\begin{table}
\caption{\label{tab:table1}Results of fitting by the Fermi distribution
with $N_f=1$, $\widetilde\rho_0$ (ch/fm$^3$), $R_0$, $t$, $r_0$ (fm),
$b$ (fm$^{-1})$, $m_\sigma$ (MeV).}
\begin{ruledtabular}
\begin{tabular}{cccccccc}
$N_q$ &$\widetilde
       \rho_0$
                  &$R_0$
                          &$b$
                                     &$t$      &$r_0$     &$m_\sigma$
&
$\eta$ \\\hline
$15$  &$0.34$     &$1.84$ &$0.51$    &$2.24$   &$0.74$    &$351$
&
$0.65$    \\
$43$  &$0.43$     &$2.19$ &$0.52$    &$2.28$   &$0.75$    &$384$
&
$0.73$    \\
$159$ &$0.46$     &$4.19$ &$0.52$    &$2.29$   &$0.77$    &$409$
&
$0.78$    \\
$303$ &$0.47$     &$5.23$ &$0.52$    &$2.29$   &$0.78$    &$417$
&
$0.795$    \\
$529$ &$0.47$     &$6.37$ &$0.52$    &$2.27$   &$0.79$    &$423$
&
$0.805$    \\
$742$ &$0.47$     &$7.15$ &$0.52$    &$2.27$   &$0.79$    &$426$
&
$0.81$ \\
\end{tabular}
\end{ruledtabular}
\end{table}

The curves plotted in the Fig. \ref{f7} and results of Table \ref{tab:table1}
allows us to conclude that the density distributions at $N_q\ge 50$
are in full agreement with the corresponding data typical for
the nuclear matter. The thicknesses of transition layers in both cases are also
similar and the coefficient $r_0$ with the factor $3^{1/3}$ included is
in full correspondence with $\widetilde r_0$.
The values of $\sigma$-meson mass in Table \ref{tab:table1} look quite reasonable as well.
However the corresponding quantities are ~strongly different at small quark numbers in the droplet.
We know from the experiments that in the nuclear matter some increase of the nuclear density is observed.
It becomes quite considerable for the Helium and is much larger than the standard nuclear
density for the Hydrogen.

Obviously, we understand the Thomas--Fermi approximation which is used for
estimating becomes hardly justified at small number of quarks,
and we should deal with the solutions of complete equation system (\ref{sys}).
However, one very encouraging hint comes from the chiral soliton model of
nucleon \cite{BB}, where it has been demonstrated  that solving this
system (\ref{sys}) the good description of nucleon and $\Delta$ can be obtained.
Then our original remark could be that the soliton solutions obtained in \cite{BB} permit an
interpretation as a ‘confluence’ of two kinks. Each of those kinks 'works'
on the restoration of chiral symmetry since the scalar field
approaches its zero value at the distance of $\sim 0.5$ fm from the kink center.
Indeed, one branch of our solution corresponds the positive value
of dynamical quark mass, and another branch presents the solution
with negative dynamical quark mass (in three-dimensional picture the pseudo-scalar
fields appear just as a phase of chiral rotation from positive to negative
value of quark mass). Such solutions develop the surface tension coefficient which is
larger in factor two than the corresponding coefficient of single
kink and as we believe signal some instability of a single kink solution.

The similar results are obtained for two flavours $N_f=2$ ($\gamma=2 N_fN_c=12$)
assuming all dynamical quark masses of SU(2) flavour multiplet are equal.
The solutions for the $\sigma$-field and density
distributions are similar to the corresponding results presented in Fig. \ref{f7}
and Fig. \ref{f7a}. The other data of fitting solutions
are shown in the following Table \ref{tab:table2}.
\begin{table}
\caption{\label{tab:table2}Results of fitting by the Fermi distribution
with $N_f=2$, $\widetilde\rho_0$ (ch/fm$^3$), $R_0$, $t$, $r_0$ (fm),
$b$ (fm$^{-1})$, $m_\sigma$ (MeV).}
\begin{ruledtabular}
\begin{tabular}{cccccccc}
$N_q$ &$\widetilde
       \rho_0$
                  &$R_0$
                          &$b$
                                     &$t$      &$r_0$     &$m_\sigma$
&
$\eta$ \\\hline
$18$  &$0.81$     &$1.56$ &$0.37$    &$1.63$   &$0.57$    &$524$
&
$0.7$   \\
$46$  &$0.9$      &$2.14$ &$0.37$    &$1.63$   &$0.6$     &$557$
&
$0.75$  \\
$169$ &$0.93$     &$3.43$ &$0.36$    &$1.6$    &$0.62$    &$586$
&
$0.79$  \\
$278$ &$0.94$     &$4.08$ &$0.36$    &$1.6$    &$0.62$    &$594$
&
$0.8$   \\
$525$ &$0.94$     &$5.04$ &$0.36$    &$1.6$    &$0.62$    &$603$
&
$0.81$  \\
$776$ &$0.94$     &$5.76$ &$0.36$    &$1.6$    &$0.63$    &$607$
&
$0.815$ \\
\end{tabular}
\end{ruledtabular}
\end{table}
As it is seen the characteristic ensemble density is approximately in factor two
larger than the density of normal nuclear matter (remember again the factor $3$).
The characteristic values of $\sigma$-meson mass
are slightly larger than for $N_f=1$ and, consequently, the thickness of the
transition layer is smaller almost in factor $1.4$. The coefficient
interrelating the mean size of droplet and the baryon (quark)
number $\widetilde r_0\sim 0.8$ is getting smaller. In principle, one can
correct (increase) the surface layer thickness and the parameter $\widetilde r_0$
by decreasing the $\sigma$-meson mass but the ensemble
density remains higher than the normal nuclear one.

\begin{figure}[!tbh]
\begin{center}
\leavevmode
\includegraphics[width=0.3\textwidth]{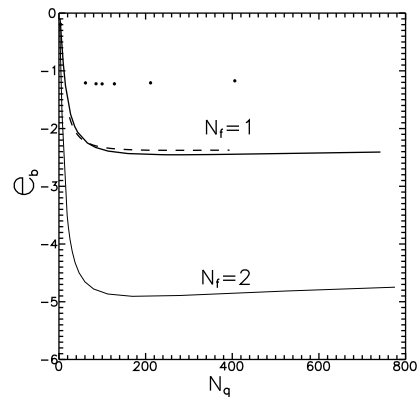}
\caption{The specific binding energy at $N_f=1$ and $N_f=2$ in MeV as a function
of quark number $N_q$.}
\label{f8}
\end{center}
\end{figure}
Fig. \ref{f8} displays the specific binding energy of ensemble. It is defined by
the expression similar to Eq. (\ref{sur}) in which the integration
over the quark droplet volume is performed. The specific energy is
normalized (compared) to the ensemble energy at the spatial infinity, i.e. in
vacuum. Actually, Fig. \ref{f8} shows several curves in the upper part
of plot which correspond the calculations with $N_f=1$. The solid line
is obtained by scanning over parameter $\eta$ and corresponds to the data
presented in Table \ref{tab:table1}. The dashed curve is calculated
at fixed $\eta=0.4$ but by scanning over parameter $\stackrel{*}M$. It is
clearly seen if the specific energy data are presented as a function of quark
number $N_q$ then the solutions, in which we are interested,
rally in the local vicinity of the curve where the maximal binding
energy -- $|{\cal E}_b|$ is reached.

The similar solution scanning can be performed over the central density
parameter $\rho_0$ in origin. The corresponding data are dotted for
a certain fixed $\stackrel{*}M$ and $\rho_0$. It is interesting to notice
that at scanning over any variable discussed a saturation property is observed
and it looks like the minimum in $e_b$ at $N_q\sim 200$--$250$.
The results for the specific binding energy as a function of particle
number are in the qualitative agreement with the corresponding experimental data.
And one may say even about the quantitative agreement if the
factor $3$ (the energy necessary to remove one baryon) is taken into account.
Another interesting fact to be mentioned is that there exist the solutions
of system (\ref{sys}) with positive specific energy. For example,
for $N_f=2$ such meta-stable solutions appear at sufficiently large
$\eta$ and with the density parameter in origin equal $\rho_0\sim
\rho_l=0.157$ ch/fm$^3$. In fact, the equation system (\ref{sys})
represents an equation of balance for the current quarks circulating between
liquid and gas phases.


As a conclusion we would like to emphasize that in the present paper we have
demonstrated how a phase transition of liquid--gas kind (with the reasonable values of parameters)
emerges in the NJL-type models. The constructed quark ensemble displays some
interesting features for the nuclear ground state (for example,
an existence of the state degenerate with the vacuum one), and the results of
our study are suggestive to speculate that the quark droplets could coexist
in equilibrium with vacuum under the normal conditions. These
droplets manifest themselves as bearing a strong resemblance to the nuclear
matter. Elaborating this idea in detail is a great challenge which will take
a lot of special efforts and we do hope to undertake them in near future.

Authors are deeply indebted K. A. Bugaev, R. N. Faustov, S. B. Gerasimov,
E.-M. Ilgenfritz, K. G. Klimenko, E. A. Kuraev, A. V. Leonidov, V. A. Petrov,
A. M. Snigirev and many other colleagues for numerous fruitful discussions.

\appendix
\section{Mean energy functional}
The free part of Hamiltonian
\begin{eqnarray}
\label{12}
&&\hspace{-0.5cm}
H_0=-\int d {\bf x}~ \bar q({\bf x})~(i {\bf \gamma}
{\bf \nabla}+im)~q({\bf x})=\nonumber\\
&&\hspace{-0.5cm}=\int \!\!\! \frac{d {\bf p}}{(2\pi)^3}|p_4|
\left[\cos \theta\left(A^+({\bf p};s)A({\bf p};s)-B({\bf p};s)B^+({\bf
p};s)\right)
\right.+\nonumber\\
&&\hspace{-0.5cm}\left.+\sin \theta \left(A^+(-{\bf p};s)B^+({\bf p};s)
+B(-{\bf p};s)A({\bf p};s)\right)\right],\nonumber
\end{eqnarray}
contributes into the mean energy as
\begin{eqnarray}
\label{13}
&&\hspace{-1.cm}\mbox{Tr}\{\xi~{\cal H}_0\}=
\int\frac{d {\bf p}}{(2\pi)^3}~
|p_4|~
(1- \cos\theta)+\nonumber\\[-.2cm]
\\ [-.25cm]
&&~~~~+\int\frac{d {\bf p}}{(2\pi)^3}~
|p_4| \cos\theta ~[n(p)+\bar n(p)]~,\nonumber
\end{eqnarray}
where ${\cal H}_0=H_0/(V2 N_c)$ is the specific energy.
Natural regularization by subtracting the free Hamiltonian $H_0$ contribution
(without pairing quarks and anti-quarks) has been done in the first term of
Eq. (\ref{13}) because in our particular situation this normalization in order
to have the ensemble energy equal zero at the pairing angle equal zero turns out
quite practical. It just explains a presence of unit in the term containing
$\cos\theta$.

The Hamiltonian part responsible for interaction,
$\bar qt^a\gamma_\mu q\bar q't^a\gamma_\nu q'$,
provides four nontrivial contributions.
The term $ \mbox{Tr}\{\rho BB^+B'B'^+ \}$ generates the following items:
$\overline{V}_{\alpha i}({\bf p},s)t^a_{ij}\gamma^\mu_{\alpha \beta}
V_{\beta j}({\bf Q},T) \overline{V}_{\gamma k}({\bf Q},T)t^b_{kl}
\gamma^\mu_{\gamma \delta}V_{\delta l}({\bf p},s)$
(the similar term but with the changes $Q,T\to Q',T'$ which generates another
primed quark current should be added) and
$-2\overline{V}({\bf Q},T)~t^a\gamma^\mu V({\bf Q}',T')
\overline{V}({\bf Q}',T')t^b\gamma^\mu V({\bf Q},T)$. Here
(as in all other following expressions) we omitted all colour and spinor
indices which are completely identical to those of previous matrix element.
The term $ \mbox{Tr}\{\rho BAA'^+B'^+\}$ generates the following nontrivial
contributions:
$\overline{V}({\bf p},s)t^a\gamma^\mu U({\bf q},t)
\overline{U}({\bf q},t)t^b\gamma^\mu V({\bf p},s)$\\
$-\overline{V}({\bf p},s)t^a\gamma^\mu U({\bf P},S)
\overline{U}({\bf P},S)t^b\gamma^\mu V({\bf p},s)$\\
$-\overline{V}({\bf Q},T)t^a\gamma^\mu U({\bf q},t)
\overline{U}({\bf q},t)t^b\gamma^\mu V({\bf Q},T)$\\
$+\overline{V}({\bf Q},T)t^a \gamma^\mu U({\bf P},S)
\overline{U}({\bf P},S)t^b\gamma^\mu V({\bf Q},T)$.
Averaging $\mbox{Tr}\{\rho AA^+A'A'^+ \}$ gives the contributions as:
$\overline{U}({\bf P},S)t^a\gamma^\mu U({\bf p},s)
\overline{U}({\bf p},s)t^b\gamma^\mu U({\bf P},S)$
(adding the similar term but with the changes $P,S\to P',S'$) and
$-2\overline{U}({\bf P},S)t^a\gamma^\mu U({\bf P}',S')
\overline{U}({\bf P}',S')t^b\gamma^\mu V({\bf P},S)$.
Another nontrivial contribution comes from averaging
$ \mbox{Tr}\{\rho A^+B^+B'A' \}$ and it has the form
$\overline{V}({\bf Q},T)t^a\gamma^\mu U({\bf P},S)
\overline{U}({\bf P},S)t^b\gamma^\mu V({\bf Q},T)$.
All other diagonal matrix elements generated by the following terms
$ \mbox{Tr}\{\rho AA^+B'B'^+\}$, $ \mbox{Tr}\{\rho BB'^+A'^+A' \}$,
do not contribute at all (their contributions equal to zero).
Similar to the calculation of matrix elements at zero temperature performed in
Ref. \cite{ff} we should carry out the integration over the Fermi sphere with
the corresponding distribution functions in the quark and anti-quark momenta
$\int^{P_F} \frac{d {\bf p}}{(2\pi)^3}$ $\to$ $\int
\frac{d {\bf p}}{(2\pi)^3}[n(p)+\bar n(p)]$
if deal with a finite temperature. All necessary formulae for the polarization
matrices which contain the traces of corresponding spinors could be found in
Refs. \cite{MZ} and \cite{ff}. Bearing in mind this fact here we present
immediately the result for mean energy density per one quark degree of freedom
as
\begin{widetext}
\begin{eqnarray}
\label{14}
w&=&\int\frac{d {\bf p}}{(2\pi)^3} |p_4| \cos\theta [n(p)+\bar n(p)]+
2 G\int \frac{d {\bf p}}{(2\pi)^3}
\sin \left(\theta-\theta_m\right)[n(p)+\bar n(p)]
\int \frac{d {\bf q}}{(2\pi)^3} \sin\left(\theta'-\theta'_m\right)I-
\nonumber\\[-.2cm]
\\ [-.25cm]
&-&G\int \frac{d {\bf p}}{(2\pi)^3}~
\sin \left(\theta-\theta_m\right)[n(p)+\bar n(p)]
\int \frac{d {\bf q}}{(2\pi)^3}~
\sin\left(\theta'-\theta'_m\right)~[n(q)+\bar n(q)]~I+\nonumber\\
&+&\int \frac{d {\bf p}}{(2\pi)^3}~|p_4|(1-\cos\theta)
-G\int \frac{d {\bf p}}{(2\pi)^3}~\sin\left(\theta-\theta_m\right)
\int \frac{d {\bf q}}{(2\pi)^3}~\sin\left(\theta'-\theta'_m\right)~
I~,\nonumber
\end{eqnarray}
\end{widetext}
(up to the constant unessential for our consideration
here)\footnote{It is interesting to notice that the existence of the angle $\theta_m$ stipulates
the discontinuity of mean energy functional mentioned above and discovered in \cite{MZ}.}.
It is quite practical to single out the colour factor in the four-fermion
coupling constant as $G=2 \widetilde G/N_c$. Performing now the following transformations
while integrating in the interaction terms
\begin{eqnarray}
&&\hspace{-0.5cm}
2 \int d {\bf p}~f\int d {\bf q} -\int d {\bf p}~ f\int d {\bf q}~f'
-\int d {\bf p}\int d {\bf q}=\nonumber\\
&&\hspace{-0.5cm}
=\int d {\bf p}~ f\int d {\bf q}~(1-f')-
\int d {\bf p}~(1-f)\int d {\bf q},\nonumber
\end{eqnarray}
and changing the variables ${\bf p}$ $\leftrightarrow$ ${\bf q}$ in the last term we obtain
\begin{eqnarray}
&&\int d {\bf p}~f\int d {\bf q}~(1-f')-
\int d {\bf p}\int d {\bf q}~(1-f')=\nonumber\\
&&
-\int d {\bf p}~(1-f)\int d {\bf q}~(1-f')~.\nonumber
\end{eqnarray}
Here the primed variables correspond to the momentum $q$. Then putting all the
terms together we come to the equation (\ref{15}).


\end{document}